# Belle II Experiment Network and Computing

excerpts from *Belle II Experiment Network Requirements Workshop Report* - DM Asner (PNNL), E Dart (ESnet), T Hara (KEK) http://www.es.net/assets/pubs_presos/Belle-II-Experiment-Network-Requirements-Workshop-v18-final.pdf [1]

The Belle experiment [2] at the KEKB accelerator [3] in Tsukuba, Japan, is a collaboration of ~400 physicists from 55 institutions across four continents. The Belle detector recorded electron-positron interactions near 10 GeV center-of-mass energy over the period 1999-2010 and has published 380 papers from these data to date. KEK has initiated an accelerator upgrade, SuperKEKB [4], designed to have instantaneous and integrated luminosity two orders of magnitude greater than those of KEKB. The new international collaboration at SuperKEKB is called Belle II [4].

The US Belle II institutions are PNNL, Carnegie Mellon University, University of Cincinnati, University of Hawaii, Indiana University, Kennesaw State University, Luther College, University of Mississippi, University of Pittsburgh, University of South Alabama, University of South Carolina, Virginia Tech, and Wayne State University. In September 2012, the US Belle II DOE Project managed by PNNL achieved the CD-1 milestone. The Belle II Collaboration includes more than 550 scientists from 23 countries – Japan, US, Australia, Austria, Canada, China, Czech, Germany, India, Italy, Korea, Malaysia, Mexico, Poland, Russia, Saudi Arabia, Slovenia, Spain, Taiwan, Thailand, Turkey, Ukraine, Vietnam.

## Computing Environment

The SuperKEKB accelerator is designed to deliver an instantaneous luminosity of $8 \times 10^{35}$ cm$^{-2}$s$^{-1}$ in 2022; thus, the Belle II experiment will collect a data sample of 50 ab$^{-1}$, corresponding to a few hundred petabytes in total data. The Belle II computing system is required to process this ever-increasing data sample without any delay to the experiment data acquisition, and to produce Monte Carlo events and physics analysis. The computing resources required for these purposes increase faster than the projected performance of CPUs and storage devices. KEK cannot be expected to provide centralized computing resources for the entire Belle II collaboration. The Belle II collaboration is expanding to more countries throughout the world, and the expectation is that collaborating institutions will contribute the Belle II computing system. Many of the Belle II institutes and universities have already started operating grid sites for the LHC experiments. After evaluating these factors, the Belle II collaboration adopted a distributed computing model based on the grid as a baseline design [5].

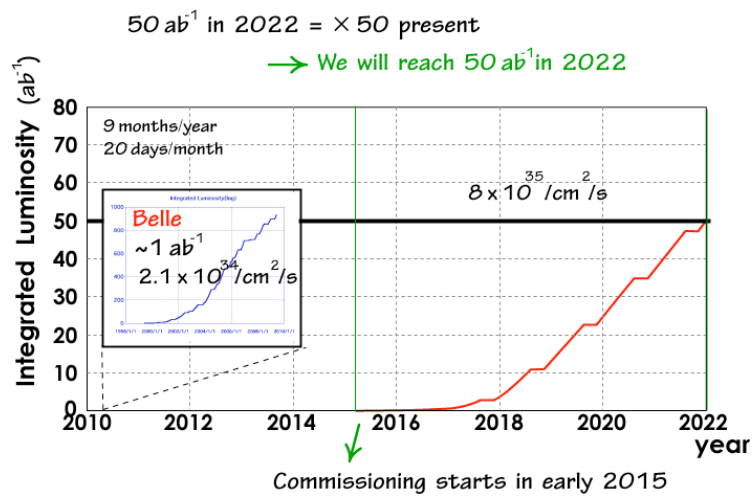

**Figure 1:** Integrated luminosity expectations profile for Belle II

*Belle II Distributed Computing Design*

The Belle II computing system has to accomplish several tasks - raw data processing, Monte Carlo (MC) sample production, physics analysis processes, and archiving the data resulting from each process. The Collaboration decided to process and store raw data at KEK. The resulting output (called "mDST"), is roughly one tenth of the size of the raw data and will be distributed to the other grid sites. This reduces unnecessary network traffic from KEK to each grid site. In other words, the raw data will not be distributed to each Tier1-level grid site. This is a simpler model than the current LHC experiments utilize. However, the Collaboration also decided to have one duplicate of the raw data at PNNL. One reason for this is that the Belle experiment faced a temporary suspension of all of analysis activities after the earthquake in 2011 because all of the processed data were stored at KEK only. Another reason is the important role of the reprocessing center outside KEK. In terms of computing, KEK's first priority is the data acquisition and processing of raw data. In parallel with this, at the early stage of the experiment (i.e. until we understand the detector performance well), the software and the detector constants must be updated often; consequently the raw data has to be reprocessed frequently. This reprocessing can be performed only at a place where the raw data is stored. If KEK were the only place to have the raw data, it would have to be done at KEK, making KEK's workload much heavier. On the other hand, with another copy of the raw data at PNNL, the reprocessing can be performed there, reducing the load on KEK. Finally, this arrangement will make the physics results available faster. Therefore, the PNNL data center, is very important for the Belle II experiment for flexibility in the raw data processing in addition to storing a backup copy of the raw data. However, the data transfer from KEK to PNNL will require a one order of magnitude higher network bandwidth between KEK/Japan and PNNL/US.

In the experience of the Belle Collaboration, MC samples at least six times larger than the beam data are required to produce precision measurements. As the MC production does not need large input files, this task can be distributed easily to the grid sites. In order to reduce peak demand, optional processing by cloud computing is planned. The output format of the MC events is the same as that of the mDST from the beam data. The MC samples will be placed on the disks on the grid site where it was produced, and at least one replica will be distributed to other grid sites.

As with the Belle II computing design, we expect users to perform analysis processes on the mDST files on the grid and to transfer the resulting lighter output (Ntuple—see Figure 2) to the local resource. The local resources will ideally be grid-enabled, but non-grid resources are explicitly included, like private clusters at institutes.

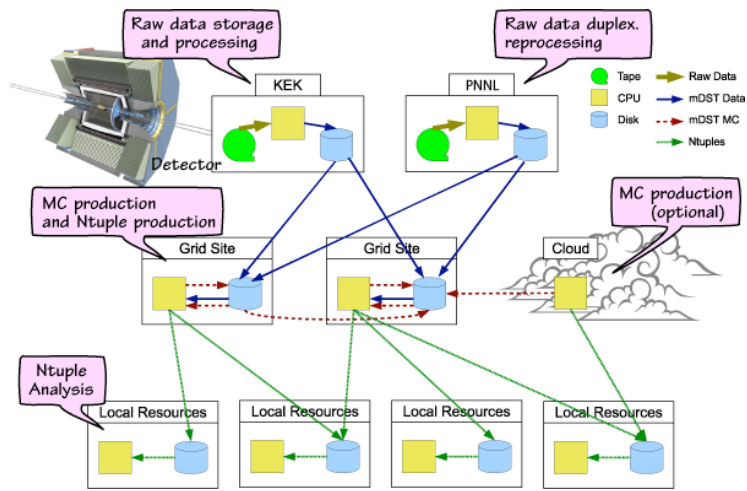

**Figure 2:** Concept of Belle II computing model

Based on the Belle II computing design and the luminosity expectations for the SuperKEKB accelerator, the total required computing resources estimated as a function of calendar year as shown in Table 1.

| Year | 2013 | 2014 | 2015 | 2016 | 2017 | 2018 | 2019 | 2020 | 2021 | 2022 |
|---|---|---|---|---|---|---|---|---|---|---|
| Tape [PB] | 2.8 | 2.8 | 2.8 | 2.8 | 19.24 | 54.43 | 103.55 | 153.89 | 204.64 | 255.39 |
| Disk [PB] | 4.00 | 4.00 | 5.00 | 8.00 | 27.98 | 79.17 | 115.68 | 153.10 | 190.82 | 228.55 |
| CPU [kHepSPEC] | 45.00 | 45.00 | 50.00 | 55.00 | 328.31 | 568.98 | 567.54 | 609.45 | 643.14 | 672.60 |

Table 1: **Total required Belle II computing resources**

KEK is expected to perform the raw data processing and storage. In addition, the tasks of MC production and user analysis are allocated in proportion to the number of Belle II PhD physicists at that site (25% is assumed for KEK in this estimation). Table 2 shows the expected computing resources at KEK. The resulting mDST files are transferred to each grid site to create at least one complete copy across the Belle II computing system, and the raw data will also be transferred to PNNL.

**KEK resources:**

| Year | 2013 | 2014 | 2015 | 2016 | 2017 | 2018 | 2019 | 2020 | 2021 |
|---|---|---|---|---|---|---|---|---|---|
| Tape [PB] | 2.80 | 2.80 | 2.80 | 2.80 | 9.62 | 27.22 | 51.77 | 76.94 | 102.3 |
| Disk [PB] | 3.00 | 3.00 | 3.00 | 3.00 | 4.57 | 12.94 | 22.44 | 32.17 | 41.98 |
| CPU [kHepSPEC] | 40.00 | 40.00 | 40.00 | 40.00 | 93.60 | 168.01 | 184.65 | 196.20 | 204.98 |
| WAN [Gbit/s] | 0.50 | 1.00 | 2.50 | 4.00 | 9.71 | 18.83 | 24.25 | 24.86 | 25.06 |

Table 2: **Required Belle II computing resources at KEK**

KEK has 2.8PB tape and 3.0PB disk storage space and 40kHepSPEC CPU power now. KEK will keep using the current computing system until 2015 summer, of which resources (The tape/disk storage and CPU) has satisfied the expected requirement.

PNNL, as a raw data storage center, plays an important role in data reprocessing. It is assumed that the reprocessing will be repeated most frequently in the first year of the data collection, then the number of reprocesses is expected to decrease as the reconstruction software matures. Finally, after four years of operation, the Collaboration must stop reprocessing activities, except in the case that more sophisticated reconstruction algorithm is invented. On the other hand, the amount of beam data will increase as the instantaneous luminosity increases. PNNL will mainly handle the reprocessing in the early stage of the experiment and evolve into a data storage role in the latter stage. PNNL will store the latest version of the mDST and the second-latest version of mDST. As with the reprocessing of the raw data, the corresponding MC samples will also be produced in proportion to the number of PhD physicists in each grid site (15% for PNNL). Another role of PNNL will be to distribute the reprocessed mDST to the Belle II grid sites. Table 3 shows the required computing resources for PNNL.

**PNNL resources:**

| Year | 2013 | 2014 | 2015 | 2016 | 2017 | 2018 | 2019 | 2020 | 2021 |
|---|---|---|---|---|---|---|---|---|---|
| Tape [PB] | 0.00 | 0.00 | 0.00 | 0.00 | 9.62 | 27.22 | 51.77 | 76.94 | 102.3 |
| Disk [PB] | 1.00 | 1.00 | 2.00 | 5.00 | 12.00 | 17.00 | 22.00 | 27.00 | 32.00 |
| CPU [kHepSPEC] | 5.00 | 5.00 | 10.00 | 15.00 | 59.11 | 95.81 | 76.58 | 82.65 | 87.63 |
| WAN [Gbit/s] | 0.50 | 1.00 | 2.50 | 4.00 | 8.65 | 15.75 | 18.82 | 19.29 | 19.44 |

Table 3: **Required Belle II resources at PNNL**

As stated above, each regional grid center is expected to perform MC production and user analysis in proportion to the number of PhD physicists assigned to that site, but this will not be the case with raw data storage. As a case study, the required computing resources for GridKa in Germany, where a 14% of the MC sample is produced, is summarized in Table 4 GridKa could become a major Belle II data center in Europe and is expected to have a full copy of the mDST for beam data and the aforementioned proportionate amount of the mDST for MC.

| Regional Center resources: | | | | | | | | | |
|---|---|---|---|---|---|---|---|---|---|
| Year | 2013 | 2014 | 2015 | 2016 | 2017 | 2018 | 2019 | 2020 | 2021 |
| Disk [PB] | 0.00 | 0.00 | 0.03 | 0.27 | 3.22 | 9.12 | 15.16 | 21.35 | 27.59 |
| CPU [kHepSPEC] | 0.00 | 0.00 | 0.68 | 4.34 | 41.40 | 71.94 | 72.21 | 77.93 | 82.63 |
| WAN [Gbit/s] | 0.00 | 0.00 | 0.04 | 0.27 | 2.65 | 4.55 | 3.86 | 3.96 | 3.99 |

Table 4: Required Belle II computing resources at 14% Regional Grid Center

The regional center will have storage space and CPU for the Belle II experiment even before taking the beam data, e.g., 2013 and 2014 for the Monte Carlo event production campaign. But this table does not include the currently available computing resources at regional centers.

## Belle II Distributed Computing Software

The Belle II experiment has adopted the grid computing model to enable the processing of the very large volume of experimental data and Monte Carlo samples that the Collaboration must analyze. In order to realize this, access to different types of computing resources, such as gLite middleware in Europe/Japan, OSG middleware in US, cloud computing and local resources are required. DIRAC, which was originally developed by LHCb and is now an independent project, can provide this environment. The DIRAC backend allows us to process jobs on the heterogeneous computing systems listed above, once the backend interface corresponding to each system is prepared. Another feature of DIRAC is the pilot job, which provides more reliable job scheduling. Furthermore, the DIRAC monitoring system is useful to check the status of jobs and resources. For metadata service software, AMGA (ARDA Metadata Grid Application), which provides efficient and scalable metadata searching is under consideration. These two servers are the core of the Belle II distributed computing system and are now operated at KEK and PNNL. This redundancy is desirable as it is preferable to have alternate servers at another site where the raw data reprocessing and the large-scale MC production are planned, e.g., PNNL.

File-level metadata catalogues (AMGA, AMI etc.) are used to improve the identification of suitable event samples. While tremendously helpful, these methods are increasingly seen as too coarse to provide an efficient analytical environment for large-scale data. The emerging concept of event-metadata-based selection and access is promising; however, current systems do not provide the necessary scalability and functionality to handle the timely analysis of extreme-scale data. Therefore, it is important that new research be performed to improve the existing event-metadata-based technology.

As previously stated, there are several tools that are used to deploy and monitor each component of the grid infrastructure. However, this monitoring information appears not to be collected presently. Developing new services/agents that collect the monitored information to optimize the overall performance of the grid virtual organization would decrease the time needed for new discoveries as well as the load on resources. Examples of improvements include network redirection/throttling, CPU and storage-load balancing, and others. The Belle II experiment provides an ideal environment in which to test new ideas, which can in turn be applied to other scientific endeavors.

To perform raw data processing, MC sample production, and user physics analysis, we developed the common software framework named "basf2." It has a software bus architecture, and a large-scale application such as MC sample production is realized by plugging in a set of a building block unit "modules"—for example the event generator, the full detector simulator, and reconstruction tools—into this framework. The modules are controlled by a "path" in which the order of the modules are defined in the steering file written in Python. The basf2 framework also supports parallel processing. For the grid environment, we developed a user interface, "gbasf2," in which the steering file is the same as basf2 jobs with some supplemental information such as the grid project name.

## Instruments and Facilities

### KEK Belle II Computing

The KEK computing system for the Belle II experiment is located at the KEK Computing Research Center. Previously, there were two computing systems at KEK, one for the Belle experiment and the other for the remaining on-going experiments and projects. However, these two have been unified into a single system since April 2012. This new system, known as "KEK Central Computing System (KEKCC)," has roughly 4,000 cores of calculation servers, 7 PB disk space and a 16-PB capacity tape storage system; it will continue operating until summer, 2015. The grid system is a part of the KEKCC, with 3,000 of 4,000 cores available for grid jobs, which corresponds to 44 kHEP-SPEC[1]. Though the total resources of KEKCC meet the requirements of the Belle II experiment until 2015, as shown in Table 2 they have to be shared with other experiments—for example J-PARC, ILC, Experimental Cosmology, Accelerator, Theory group and on-going analysis jobs for Belle. Figure 3 illustrates the configuration of the KEKCC system with the network connection.

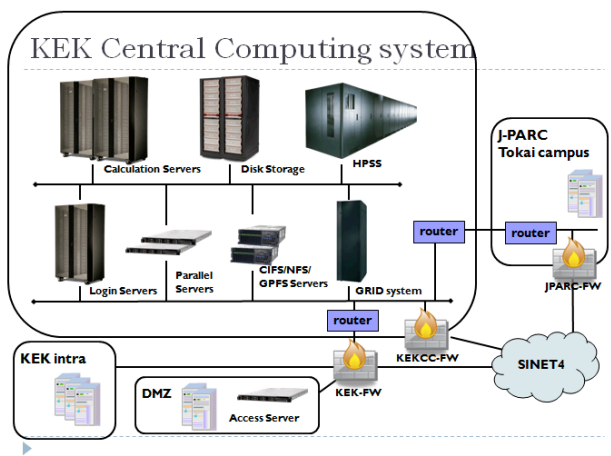

**Figure 3: Configuration of KEK Central Computing**

### PNNL Belle II Computing

Computing infrastructure for Belle II is located on PNNL's main campus in Richland, Washington USA. Compute systems and data storage will be located in the Computer Science Facility and/or the William R. Wiley Environmental Molecular Science Laboratory (EMSL) buildings. Belle II computing will be supported initially by a mixture of both dedicated and shared resources belonging to PNNL's Institutional Computing (PIC), and EMSL. Procurement of new, dedicated resources for compute and data storage are planned as the collaboration matures and funding for Belle II at PNNL grows.

---

[1] **HEP-SPEC** is the High Energy Physics-wide benchmark for measuring CPU performance, developed in 2009

At present, Belle II uses computer time and stores data on PNNL's shared computer cluster, Olympus. The collaboration also utilizes space in the EMSL High Performance Storage System (HPSS) archive and operates its own cluster of 1024 cores used for grid computing. Operational support is provided by EMSL staff in partnership with PNNL's Belle II physicists.

PNNL provides a data transfer node with 10 Gbs connectivity to both the Internet and the multiple-petabyte Lustre file system; this node provides shared data access to Olympus and to the Belle II grid nodes.

PNNL plans to meet growth in Belle II computational and data storage needs by procuring additional servers and storage annually to maximize resources per dollar as much as possible. PNNL intends to leverage this strategy to allow all Belle II data to be kept on disk. If needed, PNNL will fall back on tape storage as a cost-cutting contingency.

## Summary

To accommodate Belle II's anticipated data rates, network upgrades to allow 100 Gbps data rates will be necessary at both KEK and PNNL.

KEK is currently connected to SINET at 20 Gbps (two 10-Gbps links) at Tsukuba. SINET will be transitioning to a new network infrastructure – SINET5 – in 2016. This will be close to the time when the Belle II experiment begins production operation. It will be important for KEK, SINET, ESnet, and PNNL to collaborate closely so that SINET is aware of the needs of the Belle II experiment in time to incorporate them into the plans for SINET5 and ensure a smooth transition.

As part of the SINET5 network to be deployed in 2016. SINET is expected to provide a 100-Gbps trans-Pacific connection between Japan and Seattle. Depending on the available capacity between Japan and the United States before SINET5 is deployed, there may be a shortage of Japan-US bandwidth in 2015.

PNNL has an optical transport system that provides connectivity between PNNL and ESnet at two locations – Seattle and Boise. The current system does not have 100 Gbps capability. It is likely that upgrades to the PNNL optical transport system will be necessary to support the Belle II experiment.

In order to provide resiliency at 100 Gbps speeds, ESnet will consider adding 100 Gbps capability at its Boise location. ESnet currently has 100 Gbps capability at Seattle, the location of the primary connection between ESnet and PNNL. However, 100 Gbps connectivity for the backup connection between ESnet and PNNL at Boise should be considered.

Data and service challenge exercises have been used successfully by other experiments to verify and harden the networking, computing, and software infrastructures used in the conduct of the science. The Belle-II experiment plans several such challenges. These efforts will require coordination between the scientific, computing, and networking organizations that support the Belle-II experiment.

It is expected that PNNL will serve data to European as well as U.S. institutions. This will have implications for trans-Atlantic capacity network capacity.

Additionally, the replacement of the KEKCC computing infrastructure in 2015 may place additional demands on the trans-Pacific network infrastructure because of the desire to replicate data to the PNNL data center, both to maintain analysis continuity and to mitigate the risk of data loss during the upgrade. The KEKCC computing infrastructure is expected to be replaced again in 2018.

Undersea network cables outages are days to weeks longer than is typical for terrestrial cables. In light of this, the Belle II experiment should consider diverse paths for network connectivity between KEK and PNNL.

# Acknowledgements

This work would not have been possible without the contributions and participation of those who provided information and attended the **Belle II Experiment Network Requirements** workshop.

PNNL is operated by Battelle for the U.S. Department of Energy under contract DE-AC05-76RL01830

ESnet is funded by the U.S. Department of Energy, Office of Science, Office of Advanced Scientific Computing Research (ASCR). Vince Dattoria is the ESnet Program Manager.

ESnet is operated by Lawrence Berkeley National Laboratory, which is operated by the University of California for the U.S. Department of Energy under contract DE-AC02-05CH11231.

This work was supported by the Directors of the Office of Science, Office of Advanced Scientific Computing Research, Facilities Division, and the Office of High Energy Physics.